\documentclass[manuscript]{acmart}
\usepackage{longtable}
\usepackage{graphicx}
\usepackage{caption}
\usepackage{subcaption}
\usepackage{multirow}
\AtBeginDocument{%
  \providecommand\BibTeX{{%
    \normalfont B\kern-0.5em{\scshape i\kern-0.25em b}\kern-0.8em\TeX}}}

\copyrightyear{2022}
\acmYear{2022}
\setcopyright{acmcopyright}\acmConference[HT '22]{Proceedings of the 33rd ACM Conference on Hypertext and Social Media}{June 28-July 1, 2022}{Barcelona, Spain}
\acmBooktitle{Proceedings of the 33rd ACM Conference on Hypertext and Social Media (HT '22), June 28-July 1, 2022, Barcelona, Spain}
\acmPrice{15.00}
\acmDOI{10.1145/3511095.3531283}
\acmISBN{978-1-4503-9233-4/22/06}

\begin{document}

\title{Characterizing Vaccination Movements on YouTube in the United States and Brazil} 
\author{Marcelo Sartori Locatelli}
\authornote{Both authors contributed equally to this research.}
\email{locatellimarcelo@dcc.ufmg.br}
\orcid{0000-0002-0893-1446}
\author{Josemar Caetano}
\authornotemark[1]
\orcid{0000-0002-0522-2572}
\email{josemarcaetano@dcc.ufmg.br}
\affiliation{%
  \institution{Universidade Federal de Minas Gerais}
  \city{Belo Horizonte}
  \state{Minas Gerais}
  \country{Brazil}
}

\author{Wagner Meira Jr.}
  \orcid{0000-0002-2614-2723}
\affiliation{%
  \institution{Universidade Federal de Minas Gerais}
  \city{Belo Horizonte}
  \state{Minas Gerais}
  \country{Brazil}}
\email{meira@dcc.ufmg.br}

\author{Virgilio Almeida}
\orcid{0000-0001-6452-0361 }
\affiliation{%
  \institution{Universidade Federal de Minas Gerais}
  \city{Belo Horizonte}
  \state{Minas Gerais}
  \country{Brazil}}
  \email{virgilio@dcc.ufmg.br}

\renewcommand{\shortauthors}{Marcelo and Josemar, et al.}

\begin{abstract}
In the context of COVID-19 pandemic, social networks such as Facebook, Twitter, YouTube and Instagram stand out as important sources of information. Among those, YouTube, as the largest and most engaging online media consumption platform, has a large influence in the spread of information and misinformation, which makes it important to study how the platform deals with the problems that arise from disinformation, as well as how its users interact with different types of content. Considering that United States (USA) and Brazil (BR) are two countries with the highest COVID-19 death tolls, we asked the following question: What are the nuances of vaccination campaigns in the two countries? With that in mind, we engage in a comparative analysis of  pro and anti-vaccine movements on YouTube. We also investigate the role of YouTube in countering online vaccine misinformation in USA and BR. For this means, we monitored the removal of vaccine related content on the platform and also applied various techniques to analyze the differences in discourse and engagement in pro and anti-vaccine "comment sections". We found that American anti-vaccine content tend to lead to considerably more toxic and negative discussion than their pro-vaccine counterparts while also leading to 18\% higher user-user engagement, while Brazilian anti-vaccine content was significantly less engaging. We also found that pro-vaccine and anti-vaccine discourses are considerably different as the former is associated with conspiracy theories (e.g. ccp), misinformation and alternative medicine (e.g. hydroxychloroquine), while the latter is associated with protective measures. Finally, it was observed that YouTube content removals are still insufficient, with only approximately 16\% of the anti-vaccine content being removed by the end of the studied period, with the United States registering the highest percentage of removed anti-vaccine content(34\%) and Brazil registering the lowest(9.8\%). 
\end{abstract}

\ccsdesc[500]{Human-centered computing~Empirical studies in collaborative and social computing}
\keywords{social media, vaccination movements, textual analysis}

\maketitle

\section{Introduction}

Political polarization and disinformation are  among  the main challenges to  expand  global vaccination campaigns against the COVID-19 pandemic.  Online platforms such as Facebook, Twitter, YouTube and Instagram are social networks whose global structure shapes the flow of information and affect individual and collective behavior in issues such as vaccine,  elections,  extremism~\cite{franks2008extremism} and  racism. Social networks are also the main places where disinformation~\cite{wang2019systematic}, denialist narratives, confusion and political  disputes find a fertile ground on which to grow and spread throughout society~\cite{mitra2016understanding}. 
For these reasons, it is important to understand the dynamics of online battles between anti- and pro-vaccination groups in order to better comprehend the actions of both groups, so as to enable more effective courses of action to be taken to mitigate the consequences of such battles.
Online influence operations 
are used by different political and ideological groups to disseminate information, lies, conspiracy theories and misleading facts. United States and Brazil have  been  on the global epicenter of the  COVID-19 infodemic~\cite{nielsen2021ongoing}.  Both countries have the highest death tolls since the start of the pandemic\footnote{https://covid19.who.int/}.  According to the World Health Organization, the coronavirus infodemic undermines the global response and jeopardizes measures to control the pandemic~\cite{zarocostas2020fight}.  Different studies found that anti-vaccine groups manage to manipulate social media with effective communication strategies that capture audiences to disseminate their messages.
Commenting on online platforms have been considered one of the most popular forms of public online participation~\cite{kim2021distorting}. Comments are the way in which online collective debates take form around specific  topics  promoted by videos. They represent the relation between different groups of society with official vaccination campaigns and reflect the degree of polarization around vaccination.

Social platforms play an important role in identifying and removing online anti-vaccine misinformation to limit the consequences  it can do to a country’s vaccination movement. We monitored counter influence operations to characterize the language of comments and videos that were removed from YouTube. 

In this paper, we present two case studies that investigate the behavior of anti and pro-vaccine  groups in different  political contexts, represented by the use of social media in the United States and Brazil. In order to understand the nuances of vaccination campaigns in the two countries, we have the four following research questions, that  are explored in our analysis of the data sets we have  collected from YouTube.

\begin{itemize}
    \item What are the  characteristics that differentiate pro-vaccine and anti-vaccine online comments?
    \item What kind of language drives  engagement in online comments?
    \item What are similarities and differences between pro and anti-vaccination campaigns in United States and Brazil?
    \item What are characteristics of the pro and anti-vaccine videos removed from YouTube?
\end{itemize}

In this work we examine  characteristics  of  pro and anti-vaccine movements. 
Our analysis explores the polarization dynamics of comment sections of political and ideological YouTube channels and videos. On YouTube, we observe that both language  and engagement level are characteristics of different narratives of the ideological groups involved in  battles around vaccines.
 It has been observed by other references that frequent commenters are more political, more polarized, and more toxic than the average user of social media platforms~\cite{10.1093/joc/jqab034}. 
 Through the use of NLP tools, we  study the nature and language of comments of both pro- and anti-vaccination groups. 

Our data collection efforts resulted in a dataset with more than 8.5M comments in more than 3.0M  YouTube accounts in the two countries. We analyze this large dataset extensively: 
\begin{itemize}
\item 
We found that negative emotions are much more popular and prevalent in anti-vaccine comment sections and content.
\item
We noticed  the removal of anti-vaccine content on YouTube seems to be insufficient to stop anti-vaccine videos.
\item
We showed the contrast between the two groups: anti-vaccine discourse is associated with conspiracy theories, misinformation and alternative medicine, while  pro-vaccine discourse is related to protective measures.
\item
We found that user-user engagement is much higher on comments in removed videos, which could suggest that content deemed inappropriate leads to more discussion.
\end{itemize}

\section{Related Work}
Numerous studies have been done in the past regarding disinformation and anti-vaccine campaigns, producing countless results throughout the years. Many of those results in regards to anti-vaccination were reviewed by Hussain et al.~\cite{hussain2018anti}, who gave an overview of the movement, including the influence of technology. They mention previous work that found that 32\% of the videos on YouTube about immunization were anti-vaccine and had higher rates and more views than pro-vaccine videos~\cite{keelan2007youtube}. Another study identified that 43\% of websites found in search engines when searching for "imunization" and "vaccination" were anti-vaccine~\cite{davies2002antivaccination}.

The cross-platform nature of such campaigns was illustrated by the research conducted by~\cite{wilson2020cross}, which investigated an online cross-platform disinformation campaign targeting a rescue group operating in Syria called White Helmets. They found that anti-White Helmets accounts did more "cross-posting" from YouTube to Twitter, creating a bridge between the two platforms, directing Twitter users to a large set of videos critiquing the previously mentioned group.

Another relevant study in the context of disinformation-campaigns was conducted by~\cite{broniatowski2018weaponized}, which explores the influence of bots, Russian trolls and content polluters on vaccine discourses on Twitter. They find that these kinds of users tend to tweet about vaccination at higher rates than the average user. Additionally, they conclude that accounts posing as legitimate users create false equivalency, eroding public consensus on vaccination.
 
In this same vein,~\cite{starbird2019disinformation} examines the critical social impacts of disinformation, its extension beyond the narrow window of automated or paid actors such as bots and hired trolls—the window that tends to get the most media and research attention, and the need for more robust, cross-disciplinary treatments of information operations.
 
The competition between pro and anti-vaccination views was illustrated by~\cite{johnson2020online}, which explores the interconnected clusters across geographical locations and languages. They found that, while smaller, anti-vaccination clusters are highly entangled with undecided clusters, while pro-vaccine clusters are more peripheral. Their theoretical framework predicts the dominance of anti-vaccination views in a decade. They propose that the insights provided by the framework are used to interrupt the shift to negative views.

Focusing instead on the the role of social media companies in misinformation regarding Covid-19, Burki~\cite{burki2020online} presents distinct views by experts. While some argue that the companies' pledges to act against anti-vaccine movements were insufficient, others see issues of freedom of speech and fear that the removal of posts would turn the authors into martyrs.

By exploring user behavior in YouTube live streaming chat during Trump's 2020 presidential campaingn in the United States Liebeskind et al.~\cite{liebeskind2021analysis} find, through the use of Natural Language Processing algorithms, that most messages express sentiment, with positive sentiment outweighing the negative. They also found that heavy users tend to use more emojis and express more sentiment while using less abusive language. Finally, they concluded that sarcasm invokes engagement.

The toxicity in YouTube comment sections was studied by A. Obadimu et al.~\cite{obadimu2019identifying}, who examined comments posted on anti- and pro-NATO channels. They discovered that those on anti-NATO channels have a higher tendency to be toxic. By applying Latent Dirichlet Allocation(LDA) topic modelling, they also found that topics extracted from anti-NATO channels were more geared towards negative components and geographical locations when compared to their pro-NATO counterparts.

\section{Data Collection}

In order to understand ideological battle  on social media around vaccines, we need to look at videos, channels and  cultural contexts. We are interested in two contexts of using YouTube, represented by two different countries: the United States and Brazil. In addition to the two different countries, we focus on two classes of groups, pro-vaccination and anti-vaccination in each country.  We collected data from different YouTube channels, videos and comments  to allow us to compare distinct types of content and analyze their characteristics. For this means,
YouTube Data API v3 was used to collect data on 142 YouTube channels that contain videos related to the COVID-19 pandemic and the vaccination process. 
These channels were discovered by different means, such as keyword searches,  suggestions by the YouTube algorithm after the keyword searches, journalistic reports\footnote{These include: The Disinformation Dozen, CCDH <https://www.counterhate.com/disinformationdozen>, “Canais  na internet ganharam dinheiro com fake news sobre Covid, informa Google à CPI” (Title in English: Internet  channels made money with fake news about Covid, Google informs Parliamentary Inquiry Committee).}, inspection of YouTube links shared on Facebook’s CrowdTangle, as well as through a searchable snapshot of the Parler social network\footnote{Parler Snapshot <https://parler.adatascienti.st/>}, which includes posts, comments and users that  existed before it was shut down on January 10th. 
Channels of important and popular  political figures in the online platforms (e.g., Brazilian president Jair Bolsonaro and his  family), educational (e.g., “University of Sao Paulo channel”) or government institutions (e.g., “Ministry of Health”), several  news sources (e.g., CNN, Forbes, The New York Times), hospitals (e.g., "Hospital Israelita Albert Einstein"), well-known medical doctors (e.g., “Drauzio Varella”) or users who have a large number of subscribers or followers  (e.g., “Atila Iamarino”) were also tracked. The category (anti- or pro-vaccine) that each channel belongs to was determined manually upon inspection of the channel videos.

Data on videos posted by these channels was collected and filtered using a set of relevant keywords\footnote{shorturl.at/bpzCD}. This set of keywords is related to the COVID-19 pandemic and the vaccination process. The resulting dataset includes the video title, the channel  title, the video ID, the channel ID, the time of publication, the description of the video, tags, and topic  categories. The comments and transcript of each video that contained at least one of the keywords used were also collected. In total, 16820 videos were found to be relevant to the proposed study: 6648 anti-vaccine videos and 10172 pro-vaccine videos. The data collection period started on 11th July 2021 and ended on 3rd October 2021.

A total of 8.540.234 comments were collected, of which 172.578 had at least one hashtag and 107.429 had a link to at least one of the following websites
: YouTube, Facebook, Instagram, Twitter, Reddit, and Gab. Table \ref{tab:yt-desc} shows basic statistics grouped by Channel Category and Channel Country for the dataset with YouTube comments. Table \ref{tab:links_yt} shows the distribution of links to relevant social networks within the database.

\begin{table*}[!ht]
\centering
\resizebox{\textwidth}{!}{%
\begin{tabular}{@{}ccrrrr@{}}
\toprule
\multicolumn{1}{l}{\textbf{Channel Category}} & \multicolumn{1}{l}{\textbf{Channel Country}} & \multicolumn{1}{l}{\textbf{Accounts Number}} & \multicolumn{1}{l}{\textbf{Total Comments}} & \multicolumn{1}{l}{\textbf{Comments per Account}} & \multicolumn{1}{l}{\textbf{\% incl URL}} \\ \midrule
Anti-Vaccine                       &BR                                                      & 539.678                                      & 2.308.712                                   & 4,28                                              & 21,54\%                                  \\
Pro-Vaccine                         &BR                                                     & 350.774                                      & 649.644                                     & 1,85                                              & 22,33\%                                                    \\
Anti-Vaccine                                  & USA                                           & 280.979                                      & 584.844                                     & 2,08                                              & 13,50\%                                   \\
Pro-Vaccine                                   & USA                                           & 1.233.565                                    & 3.017.758                                   & 2,45                                              & 12,99\%                                 \\
Anti-Vaccine                                  & OTHER                                        & 248.463                                      & 709.947                                     & 2,86                                              & 14,57\%                                  \\
Pro-Vaccine                                 & OTHER                                        & 437.707                                      & 1.261.304                                   & 2,88                                              & 18,14\%                                  \\ \bottomrule
\end{tabular}%
}
\caption{Basic Statistics by Channel Category and Country for the YouTube Dataset. \textbf{\% incl URL} represents the percentage of the total number of comments that contain URLs. While building this table, we discarded 8.025 comments that have neither country nor category.}
\label{tab:yt-desc}
\end{table*}

\begin{table}[H]
\centering
{%
\begin{tabular}{@{}ccrrrrr@{}}
\toprule
\multicolumn{1}{c}{\textbf{URL}}& \multicolumn{1}{l}{\textbf{Brazil}} & \multicolumn{1}{l}{\textbf{United States}} & \multicolumn{1}{l}{\textbf{Others}}\\ \midrule
\textbf{youtu.be}   &9.485                                                                         & 31.520                                      & 23.601                                                     \\
\textbf{youtube.com}  & 4.432                                                                         & 18.976                                    & 15.602                                                    \\
\textbf{facebook.com}   &759                                                                    & 1.619                                      & 1.382                                                      \\
\textbf{twitter.com}     & 161                                                                & 979                                      & 1.023                                                      \\
\textbf{instagram.com}  &191                                                                     & 255                                      & 181                                                    \\
\textbf{reddit.com}    &3                                                                   & 135                                      & 100
\\
\textbf{gab.com} &7                                                                      & 255                                      & 181                                   
\\ \bottomrule
\textbf{Total}&15.038 & 53.739 & 42.070
\\ \bottomrule
\end{tabular}%
}
\caption{{Number of YouTube comments with links to websites by country.}}
\label{tab:links_yt}
\end{table}

\section {Methods and Tools}

We use a four-dimension approach to analyze the content of  comments, and videos. Through our analyses, we evaluate  (a) engagement, (b) tone of the language, (c) toxicity and (d) lexical analysis. We also developed a tool to monitor counter misinformation operations, i.e videos and channels removal.

\subsection{Engagement}

User-user engagement, such as the number of replies to comments, are useful measures for understanding how users interact with different kinds of content. This may be useful in the context of influence operations in social media since, by finding how the engagement varies among videos/posts, it is possible to evaluate whether or not the comment section seems to be influenced by the nature of the content.

Each social media platform allows their users to interact in different ways, for example, while Twitter and Gab allow retweeting/reblogging, liking posts, replying to posts and many other forms of interaction, YouTube comments only track replies and likes. As such, the engagement metric used for each platform has to be calculated differently to account for such diversity.

YouTube user-user engagement has previously been explored in literature~\cite{ksiazek2016user}, where it was measured as the total number of replies in a given comment divided by the number of comments in the video where it was posted:
\begin{equation}
\label{eq:engagement}
    \beta = \dfrac{\operatorname{Total\ number\ of\ replies}}{\operatorname{Number\ of\ comments\ in\ video}}, \end{equation}
where $\beta$ denotes the reply-based engagement.

When looking at user-user engagement metrics, it is important to consider the popularity of the analyzed videos, as previous studies~\cite{ksiazek2016user} show a negative relationship between video popularity and this metric. 

Besides, for US videos, specifically, VADER~\cite{hutto2014vader} sentiment analysis,a tool which is tailored for English text and maps lexical features to emotion intensities known as sentiment scores, was used to determine comments' sentiment. By using the sentiment metrics for each comment, we wanted to explore if the sentiment of a comment interfered with its engagement and whether the category(anti or pro-vaccine) of the video in which the comment was posted lead to comments of a given sentiment being more or less popular. As previous studies have shown that in other social networks such as Twitter~\cite{featherstone2020exploring}, negative sentiments lead to higher engagement regardless of the nature of the content, it was interesting to explore if such tendencies were also true for YouTube.

\subsection{Tone of comments}
We want to understand whether variations regarding the tone of the comments vary as a function of the video nature (i.e., pro- or anti-vaccine), as well as whether the tone is different when people mention other platforms in an effort to increase the reach of the comment. 

\subsubsection{Valence}
 We look at how words are associated with each group using the measure of political valence proposed in~\cite{Truthy_icwsm2011politics}. We adapt political valence to the context of this study as \textit{a measure that encodes the relative prominence of a term among Pro-Vaccine and Anti-Vaccine users}. Valence \emph{V} of a term \emph{t} is defined as: 
\begin{equation}
\label{eq:valence}
    V(t) = 2 * \frac{\frac{N(t,A)}{N(A)}}{\frac{N(t,P)}{N(P)} + \frac{N(t,A)}{N(A)}} -1
\end{equation}
where \emph{N(t,P)} and \emph{N(t,A)} are the counts of occurrences of a term \emph{t}, and \emph{N(P)} and \emph{N(A)} is the number of total terms in the Pro-Vaccine and Anti-Vaccine texts (i.e., comments), respectively. This equation bounds the valence between -1, for terms only used in Pro-Vaccine texts, and +1, for terms appearing only in Anti-Vaccine texts. 

The valence metric was then used to analyze the usage of words in comments from Anti-Vaccine and Pro-Vaccine channels during 2020 and 2021, seeking to capture the changes in the tone of comments from one year to another.

\subsubsection{Toxicity}
 
 The toxicity of comments was evaluated using the \texttt{Perspective API}~\footnote{https://www.perspectiveapi.com/} tool. We worked with the definition of toxicity used by the tool to identify whether a text (i.e., comment) could be perceived as toxic.  Toxicity refers  to ``a rude, disrespectful, or unreasonable comment that is likely to make you leave a discussion''. Perspective is a tool developed by Google that leverages deep learning to estimate the toxicity score of a text. Therefore, we can use such score to measure how much toxicity on average is published in comments on American and Brazilian channels.

\subsection{Lexical Categories Analysis}

\begin{figure*}
        \centering
        \begin{subfigure}[b]{0.475\textwidth}
            \centering
            \includegraphics[width=\textwidth]{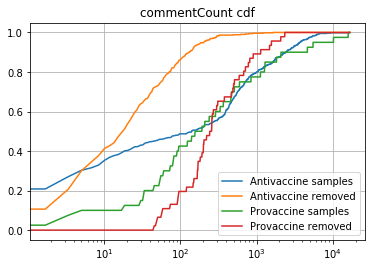}
            \caption[Comment count ECDF]%
            {{\small Comment count ECDF}}    
            \label{fig:Comment count ECDF}
        \end{subfigure}
        \hfill
        \begin{subfigure}[b]{0.475\textwidth}  
            \centering 
            \includegraphics[width=\textwidth]{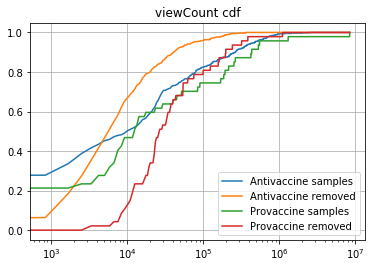}
            \caption[]%
            {{\small View count ECDF}}    
            \label{fig: View count ECDF}
        \end{subfigure}
        \vskip\baselineskip
        \begin{subfigure}[b]{0.475\textwidth}   
            \centering 
            \includegraphics[width=\textwidth]{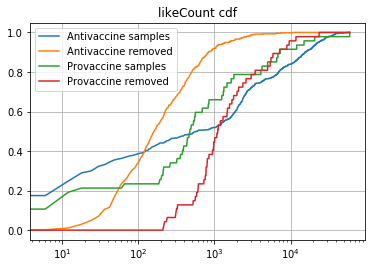}
            \caption[]%
            {{\small  Like count ECDF}}    
            \label{fig: Like count ECDF}
        \end{subfigure}
        \hfill
        \begin{subfigure}[b]{0.475\textwidth}   
            \centering 
            \includegraphics[width=\textwidth]{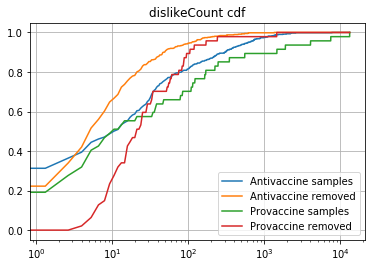}
            \caption[]%
            {{\small  Dislike count ECDF}}    
            \label{fig: Dislike count ECDF}
        \end{subfigure}
        \caption[Video Statistics ECDFs]
        {\small Video Statistics ECDFs} 
        \label{fig:Video_Statistics_ECDFs}
    \end{figure*}

Given the considerable number of video removals, it became interesting to ascertain  why the removal of these videos occurred and, thus, it was important to explore their statistics when compared to other videos of the same category that are still online. For this, a sample with the same number of videos was drawn from the dataset and their statistics were collected for comparison. The ECDF distributions in Figure \ref{fig:Video_Statistics_ECDFs} show the distribution of video statistics for the sample and removed videos.

To further compare the removed videos and the sample, it was interesting to study the difference in emotions present in their transcripts and comment sections. For this,the Empath~\cite{fast2016Empath} tool was used, as it is useful for analyzing text across lexical categories and also generating new lexical categories to use for an analysis. For this analysis, the following categories were used among the available Empath categories: 

•	positive: 'joy', 'love', 'politeness'.

•	negative: 'aggression', 'anger', 'disgust',  'hate', 'kill', 'nervousness', 'pain', 'rage', 'sadness', 'suffering', 'terrorism', 'violence'.

Ottoni et al.~\cite{ottoni2018analyzing} proposed an analysis of right-wing YouTubers using Empath which could be used for our analysis, thus, following their proposed methodology, the following steps as well as the textual processing steps of lemmatization were performed. First, the Empath for transcripts and comments were obtained, then, the mean Empath for comments from each video was obtained. After that, the mean Empath for each channel was calculated, followed by the cosine similarity between comment and transcript Empath for each video.The similarity between a video's comments and its transcript is defined as
\begin{equation}
\label{eq:similarity}
    S_v=\cos(E_{v,transcript},\overline{E}_{v,comments}),
\end{equation}
where $S_v$ denotes the similarity between the comments and transcript of video v, $E_{v,transcript}$ denotes the Empath metrics for the video transcript and $\overline{E}_{v,comments}$ denotes the mean Empath metrics for the video comments.




By following the proposed steps using these relevant Empath categories, it is possible to obtain information on how much each comment and video is related to a given category. This information is useful to determine which categories prevail on a given kind of content and can be used to correlate video and comments content for each Empath category.

As Empath currently does not natively support Portuguese, this analysis was only done for American videos.

.

\subsection{Counter misinformation operations}

It is well-known that while YouTube is an important source of vaccine-related information to many people, it is also plagued by misinformation~\cite{donzelli2018misinformation}. The platform is aware of this issue and follows four principles in an attempt to mitigate the impact of fake news in its environment: 
(i) rewarding trusted content creators, 
(ii) removing content that violates YouTube policies, 
(iii) reducing recommendations of borderline content, and 
(iv) raising up authoritative sources for news and information sources\footnote{https://www.youtube.com/howyoutubeworks/our-commitments/fighting-misinformation/}. However, the measures taken in accordance to the principles above have been shown to be insufficient~\cite{hussein2020measuring}.

In this study, we analyze the content removal in YouTube by the means of tracking the status of the videos on the database mentioned above. This was achieved using the YouTube Data API v3 on a daily basis
to detect video take downs, be it by the user or by the platform. The period during which video removals were tracked ranges from May 30,2021 to October 3,2021.

\section{Results}

\subsection{Engagement}
Tables \ref{tab:Views_categoryUS} and \ref{tab:Views_categoryBR} show the mean YouTube video views in the United States and in Brazil respectively. As mentioned, it is important to know the video popularity before looking at engagement metrics, since it is expected that lower popularity leads to higher engagement. The data from these tables will be discussed in the following sections.

\begin{table}[!ht]
\centering
\begin{tabular}{lll}
\toprule
\multicolumn{1}{l}{\textbf{}}& \multicolumn{1}{l}{\textbf{Absolute}} & \multicolumn{1}{l}{\textbf{Relative}}\\ \midrule
All videos&	143139.5&	1.0\\
Anti-vaccine&	57576.5&	0.40\\
Pro-vaccine&	188843.2&	1.31\\
All Removed videos&	20095.2&	0.14\\
Removed Anti-vaccine&	16256.9&	0.11\\
Removed Pro-vaccine&	84449.3&	0.58\\  
\bottomrule
\end{tabular}
\caption{Mean YouTube views per category on American videos}
\label{tab:Views_categoryUS}
\end{table}

\begin{table}[!ht]
\centering
\begin{tabular}{lll}
\toprule
\multicolumn{1}{l}{\textbf{}}& \multicolumn{1}{l}{\textbf{Absolute}} & \multicolumn{1}{l}{\textbf{Relative}}\\ \midrule
All videos&	71190.1&	1.0\\
Anti-vaccine&	82902.3&	1.16\\
Pro-vaccine&	60948.1&	0.85\\
All Removed videos&	 41124.0& 0.57\\
Removed Anti-vaccine&	41124.0&	0.57\\
Removed Pro-vaccine*&	-&	-\\ \hline

 \multicolumn{3}{l}{*There were no removed pro-vaccine videos in Brazil}\\
\bottomrule
\end{tabular}
\caption{Mean YouTube video views per category on Brazilian videos}
\label{tab:Views_categoryBR}
\end{table}

\subsubsection{United States}
Table \ref{table:engagement_category} shows the relative mean engagement with YouTube comments for each category and sentiment. The numbers were normalized by dividing all values by the mean engagement of all the comments in American videos. It is  important to acknowledge that the comments from removed videos were obtained before the video was taken down. 

\begin{table*}[!ht]
\centering
\resizebox{\textwidth}{!}{%
\begin{tabular}{llllll}
\toprule
\multicolumn{1}{l}{\textbf{}}& \multicolumn{1}{l}{\textbf{General}} & \multicolumn{1}{l}{\textbf{Anti-vaccine}} & \multicolumn{1}{l}{\textbf{Pro-vaccine}} & \multicolumn{1}{l}{\textbf{Removed anti-vaccine}} & \multicolumn{1}{l}{\textbf{Removed pro-vaccine}}\\ \midrule
All comments &	1.0 &	1.13 &	0.95 &	8.11 &	3.82 \\ 
Comments w/links &	0.59 &	0.82 &	0.55 &	4.35 &	3.90 \\
Comments w/hashtags	& 0.72	& 0.95 & 	0.68 &	16.96 &	2.32 \\\hline
\multicolumn{5}{c}{\textbf{Negative sentiment}} \\ \hline
All comments  &	0.95 &	1.24 &	0.86 &	9.64 &	4.66 \\
Comments w/links &	0.92 &	1.38 &	0.84 &	3.44 &	3.37 \\
Comments w/hashtags &	1.08 &	1.24 &	1.04 &	11.79 &	2.79 \\\hline
\multicolumn{5}{c}{\textbf{Positive sentiment}} \\ \hline
All comments &	1.15 &	1.14 &	1.16 & 	7.51 &	3.77 \\
Comments w/links &	0.91 &	0.91 &	0.90 &9.38 &	6.99 \\
Comments w/hashtags &	0.85 &	1.42 &	0.73 &	32.37 &	4.50\\ \hline
\multicolumn{5}{l}{*The mean engagement of all comments is 0.0006061521773068392}\\ \bottomrule
\end{tabular}%
}
\caption{Mean relative user-user engagement of comments per category of American YouTube videos}
\label{table:engagement_category}
\end{table*}

It can be seen that there is significantly higher engagement for removed videos irrespective of the sentiment of the comment. It is also interesting that while, in general, negative pro-vaccine comments show lower engagements, the opposite happens with positive sentiments, which may suggest that optimism is a more popular, or at least more discussed, view in this kind of content. In anti-vaccine content, however, negative sentiments are slightly more engaged with, which could suggest that pessimism causes more discussion in this category. Additionally, the fact that both positive and negative sentiments have greater mean engagement than the mean for all anti-vaccine comments is explained by neutral comments having lower mean engagement. 

Another interesting aspect to consider is the much higher user-user engagement in the removed videos, which could be an indicative that content deemed inappropriate by YouTube (or by the user who decided to remove their own video), sparks more discussion.

Additionally, comments with links to other social network websites or other YouTube videos have shown, in most categories, comparable mean engagement to the mean for that category. But, when considering the whole database, they have shown considerably lower mean engagement. This could suggest that users in YouTube comment sections, in general, are less interested in links to other websites/videos than in purely conversational comments.

When contrasting the relative popularity of content belonging to a given category and its user-user engagement, it can be seen that, while the expectation would be for less popular(on average) videos to have higher engagement, it can be seen that this does not always happen, for example, despite being more popular on average, comments in removed pro-vaccine videos have much lower mean engagement when compared to comments in anti-vaccine videos.

\subsubsection{Brazil}
Table \ref{table:engagement_br} shows the relative mean engagement with YouTube comments for each category for Brazil. The numbers were obtained by dividing all values by the mean engagement of all the comments in Brazilian videos.
\begin{table*}[!ht]
\centering
\resizebox{\textwidth}{!}{%
\begin{tabular}{llllll}
\toprule
                        & \textbf{General}     & \textbf{Anti-vaccine}    & \textbf{Pro-vaccine}    & \textbf{Removed anti-vaccine}    & \textbf{Removed pro-vaccine**}    \\ \hline
All comments            & 1.0                  & 0.48                     & 1.32                    & 0.69                             & -                                 \\
Comments w/links        & 0.83                 & 0.18                     & 3.45                    & 0.48                             & -                                 \\
Comments w/hashtags     & 0.25                 & 0.14                     & 1.11                    & 0.27                             & -                                 \\ \hline
\multicolumn{6}{l}{\begin{tabular}[c]{@{}l@{}}*The mean engagement of all comments is 0.0008826060131351426\\ **There were no removed pro-vaccine videos in Brazil\end{tabular}} \\ \hline
\end{tabular}
}
\caption{Mean relative user-user engagement of comments per category of Brazilian YouTube videos}
\label{table:engagement_br}
\end{table*}

It can be seen that, for Brazil, anti-vaccine content had much lower mean user-user engagement when compared with pro-vaccine, which is the opposite of what happened in the case of the United States. However, similarly to the United States, comments with links and hashtags in general had relatively lower engagement, which could suggest that YouTube users are more interested in purely conversational comments.

Unlike in the United States, removed Brazilian videos did not show extreme user-user engagement, however, they did show greater engagement than the videos of the same category that were not removed.

\subsection{Tone of comments}
\subsubsection{Valence}
\begin{figure*}[ht]
    \centering
    \includegraphics[width=\textwidth]{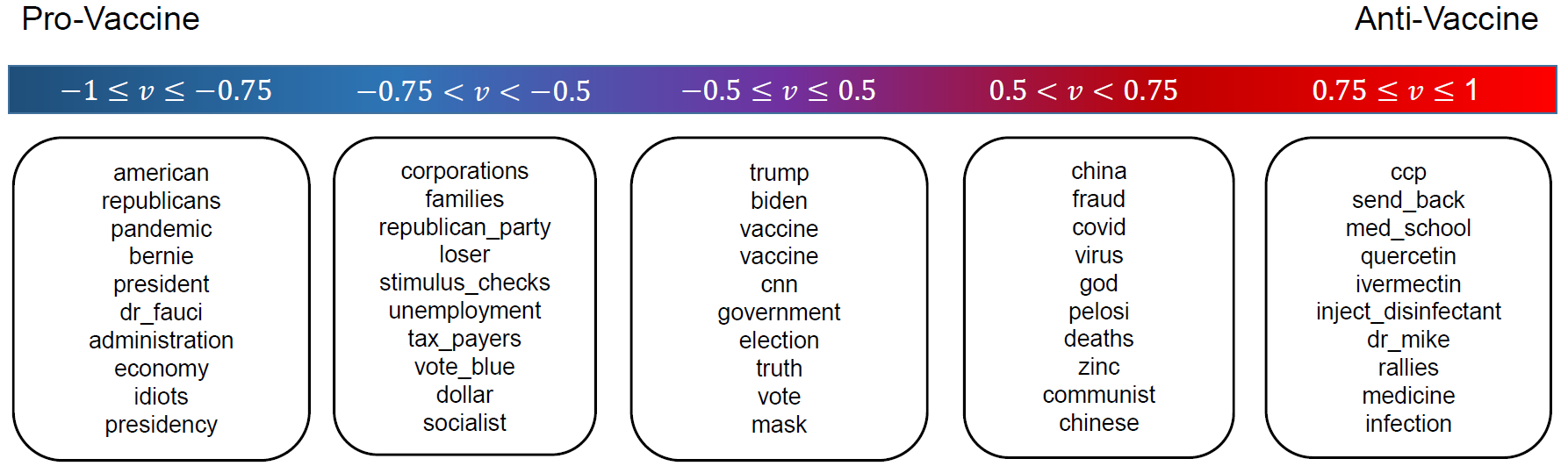}
\caption{Valence of words used in 2020 and 2021 in comments from US.}
    \label{fig:valence_words_all_us}
\end{figure*}

\begin{figure*}[ht]
    \centering
    \includegraphics[width=\textwidth]{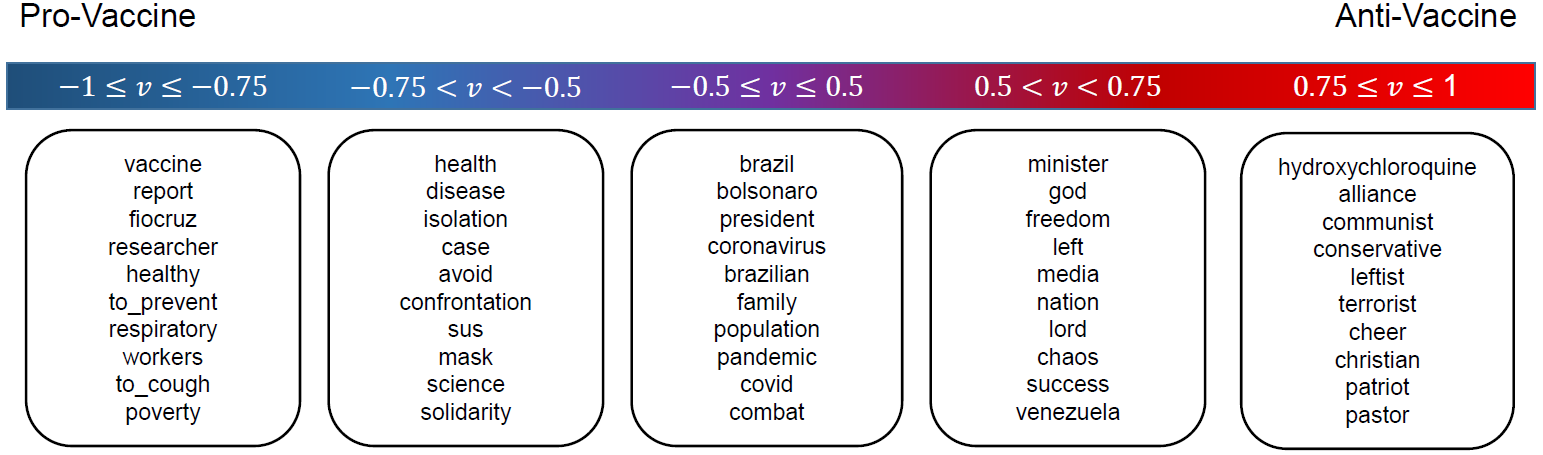}
\caption{Valence of words used in 2020 and 2021 in comments from Brazil.}
    \label{fig:valence_words_all_br}
\end{figure*}

Figure~\ref{fig:valence_words_all_us} and Figure~\ref{fig:valence_words_all_br} shows valence score of words used, from April 2020 to June 2021, in US and Brazilian comments ~\footnote{The words in the figure are  translated from Portuguese to English.}, respectively. We observe that, in both countries, anti-vaccine discourse is more associated with conspiracy theories(e.g., communist, ccp, chinese), misinformation and alternative medicine (e.g., inject\_disinfectant, quercetin, ivermectin, hydroxychloroquine). Whereas pro-vaccine discourse is associated with election and protective measures (e.g. president, administration, mask, healthy).

\subsubsection{Toxicity}
\begin{figure*}
        \centering
        \begin{subfigure}[b]{0.475\textwidth}
            \centering
    \includegraphics[width=\columnwidth]{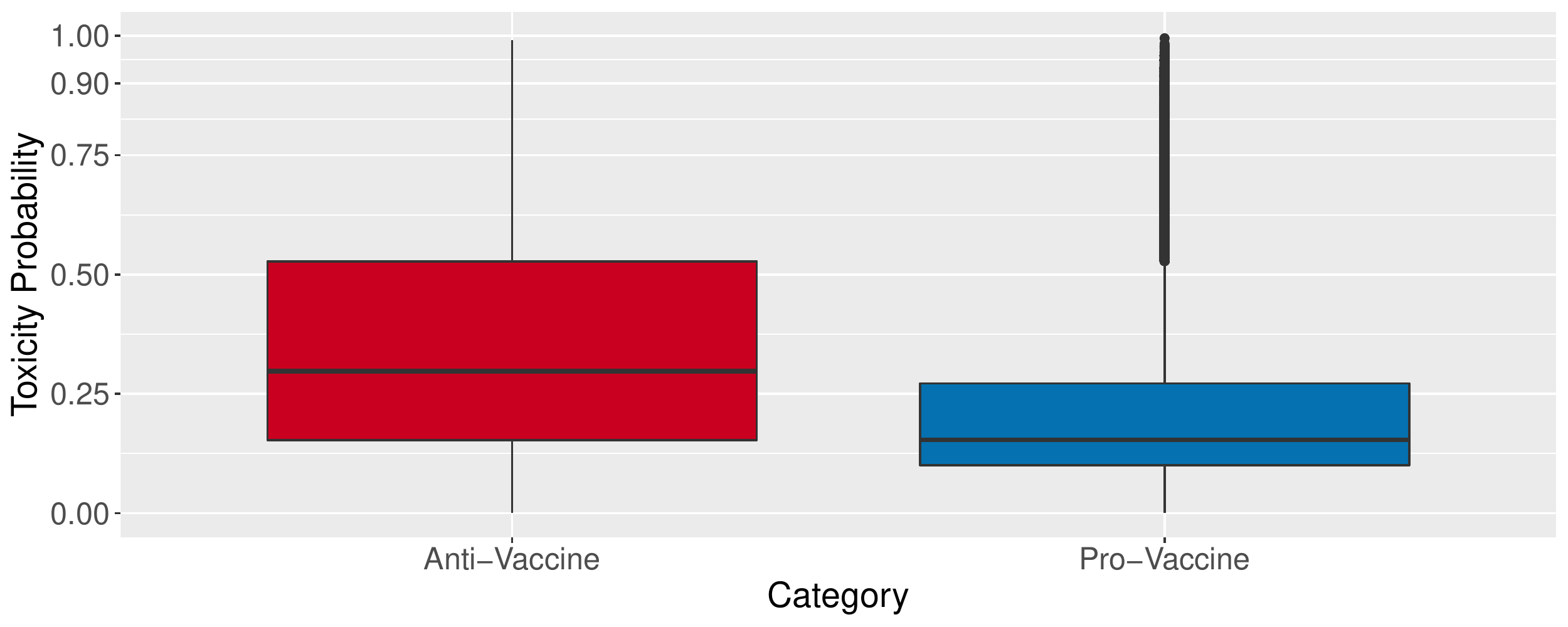}
\caption{Toxicity probability of YouTube comments from US.}
    \label{fig:toxicity_probability}
        \end{subfigure}
        \hfill
        \begin{subfigure}[b]{0.475\textwidth}  
            \centering 
    \includegraphics[width=\columnwidth]{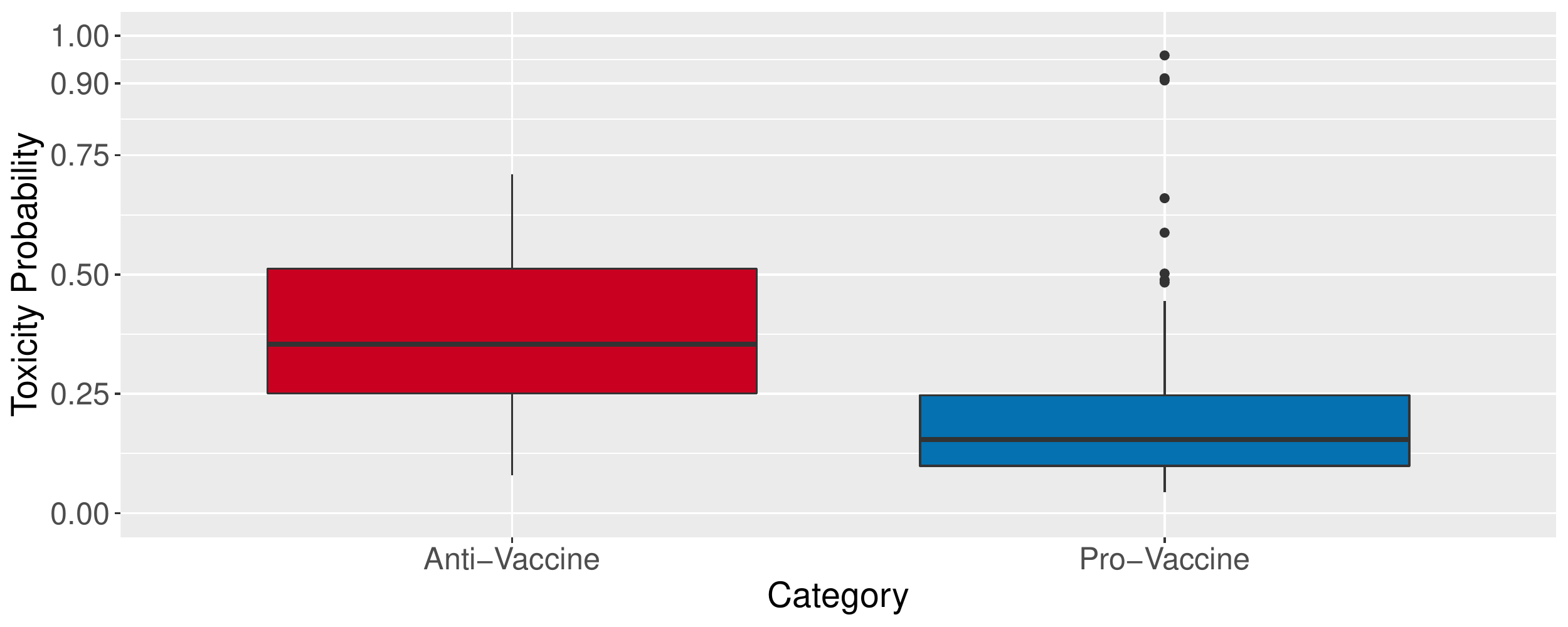}
\caption{Toxicity probability of YouTube comments from Brazil.}
    \label{fig:toxicity_probability_br}
        \end{subfigure}
\caption{Toxicity probability of YouTube comments.}
\end{figure*}

Figure~\ref{fig:toxicity_probability} and Figure~\ref{fig:toxicity_probability_br} show the toxicity probability of US comments and Brazilian comments, respectively. We observe that, in both countries, comments on Anti-Vaccine channels have more toxicity probability than those on Pro-Vaccine channels. For instance, in the United States, 75\% of comments on Anti-Vaccine channels have toxicity probability between 0.125 and 0.55, whereas on the Pro-Vaccine channels, 75\% of comments have toxicity probability between 0.09 and 0.27. In Brazil, we observe a similar characteristic. However, the majority of comments (75\%) published in Pro-Vaccine are even less toxic than the comments published in Pro-Vaccine in the United States.


\subsection{Lexical Categories Analysis}
\begin{figure*}[!ht]
    \centering
    \includegraphics[width=\textwidth]{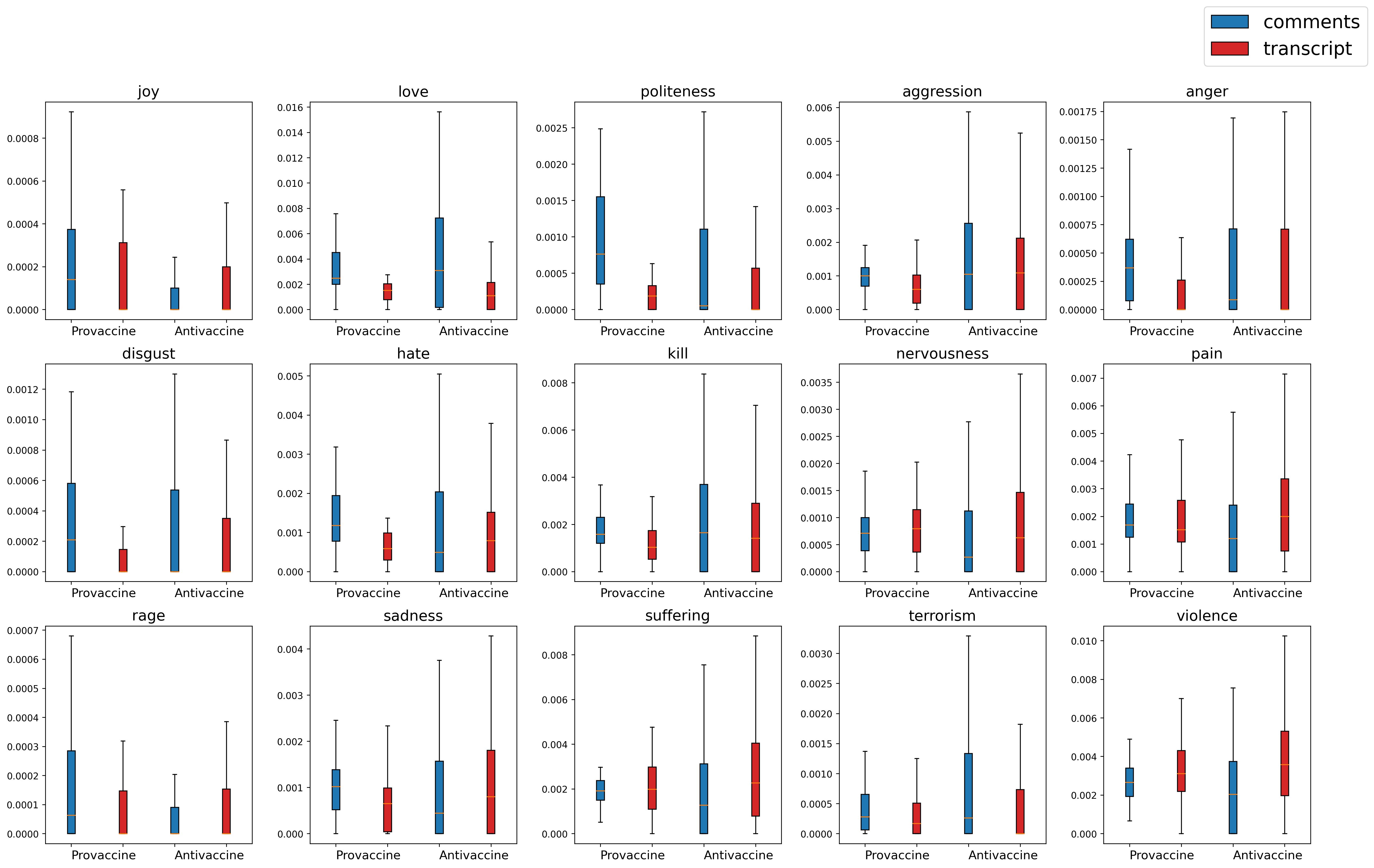}
    \caption{Empath for removed videos comments and transcripts. The y axis represents the mean Empath metric.}
    \label{fig:Empathremoved}
\end{figure*}

\begin{figure*}[!ht]
    \centering
    \includegraphics[width=\textwidth]{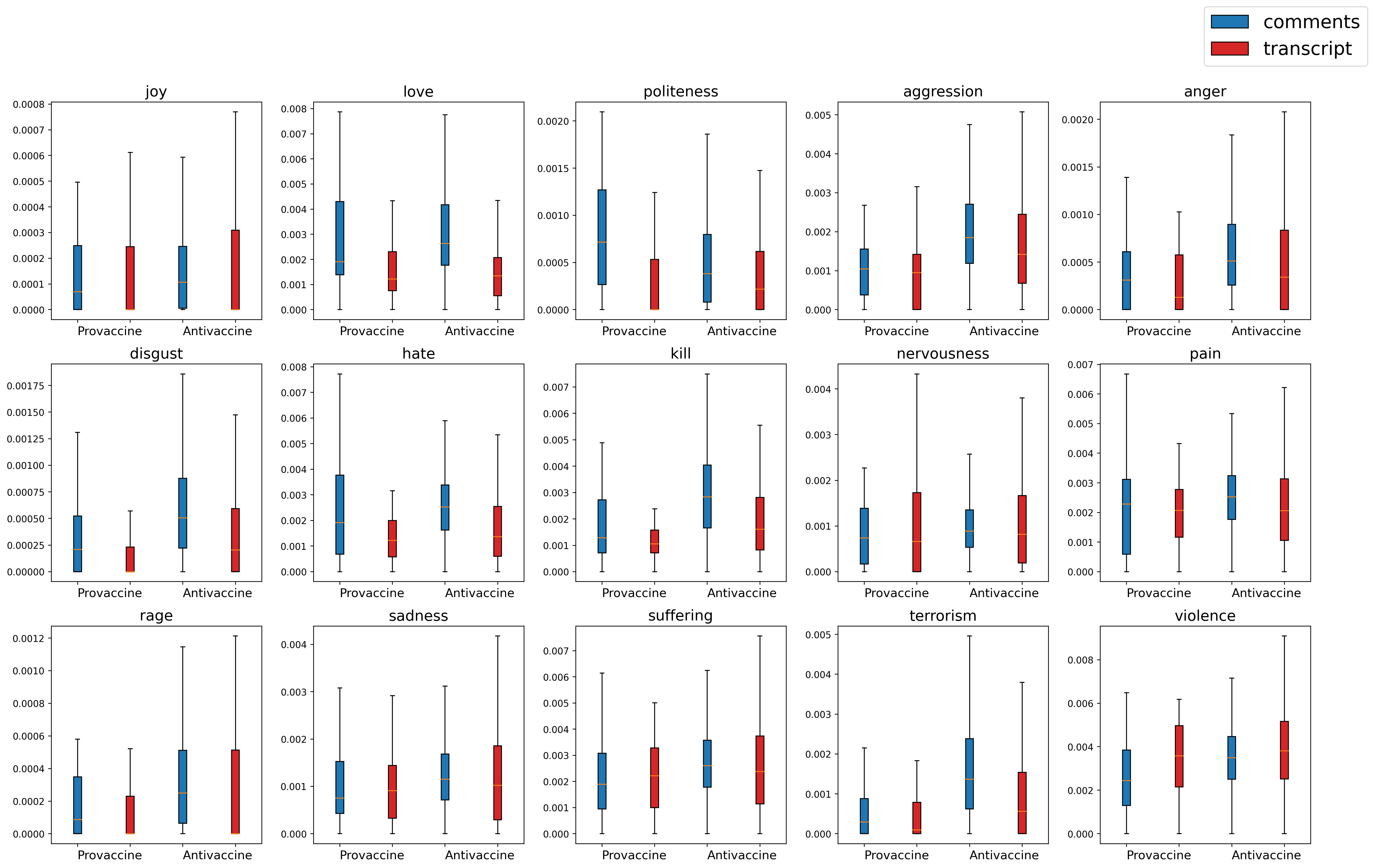}
    \caption{Empath for sample videos comments and transcripts.The y axis represents the mean Empath metric.}
    \label{fig:Empathsample}
\end{figure*}

To compare the different video categories, for both removed as well as the samples, the previously mentioned mean Empath metric for each video was used to generate Figures \ref{fig:Empathremoved} and \ref{fig:Empathsample}.

From them, it can be seen that in all categories, comments tend to have greater metrics in positive categories, while transcripts do not seem to show higher scores in negative categories in general.  Additionally, transcripts from videos that are still online seem to have greater positive scores than their removed counterparts.

Besides that, it can be observed that surprisingly, comments in the sampled non-removed pro-vaccine videos are more negative than those from removed videos, which could be a consequence of the low sample size. However, comments on anti-vaccine videos have shown greater negative scores in all categories when compared to the sample, with the exception of the “rage” category. Unsurprisingly, in most categories, especially for the removed videos, anti-vaccine content has greater negative scores.

\begin{figure*}[!ht]
    \centering
    \includegraphics[width=\textwidth]{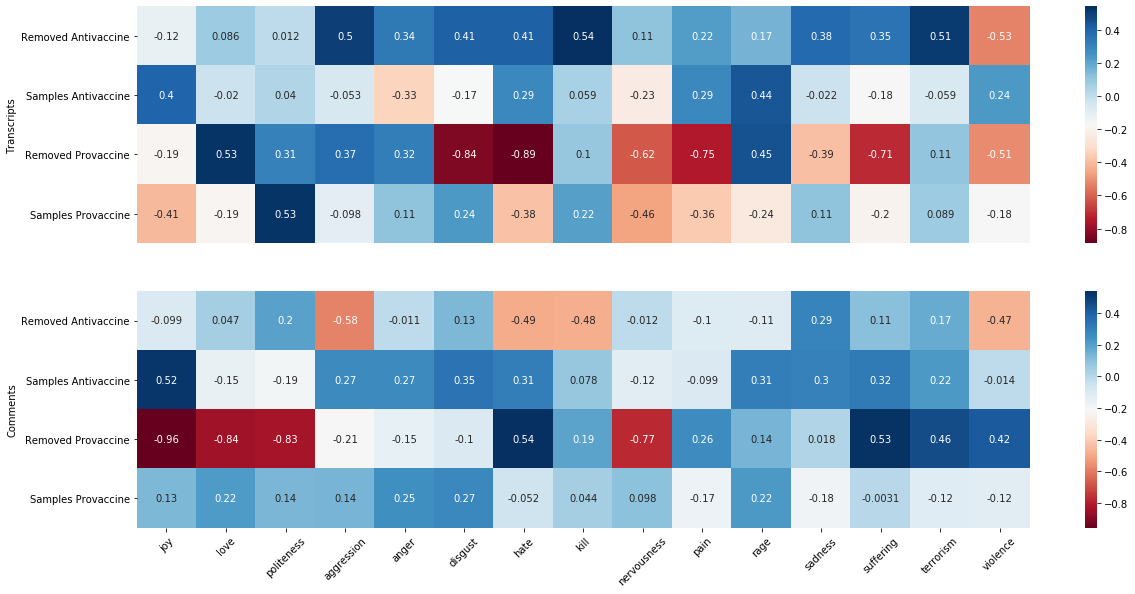}
    \caption{Correlation between similarities and Empath categories}
    \label{fig:correlationEmpath}
\end{figure*}

Figure \ref{fig:correlationEmpath} presents a matrix in which each element is the correlation between average similarities and an Empath category. From it, it can be seen that pro vaccine videos with less fraction of negative emotions on their transcripts have high similarity between caption and comments.  Anti-vaccine transcripts, however, showed positive correlations with negative Empath categories, suggesting that a higher fraction of negative emotions correlates to a high similarity, in other words, whenever content creator and commenters' views are aligned, anti-vaccine videos show more negative emotions, while pro-vaccine videos show less negative emotions. Both of these statements are especially true for removed content, as the correlations in absolute value, are higher for them.
 
Looking at the comments it can be seen that removed pro vaccine comment sections have low correlation with positive emotions. Furthermore, removed anti-vaccine comments also show low correlations for some of the negative Empath categories.

\subsection{Counter misinformation operations}

During the data collection period, a total of 1141 videos were removed, of which 648 are from the United States, 306 are from Brazil, and 187 are from other countries. It is important to highlight that while most removed videos are anti-vaccine, 47 pro-vaccine videos were removed in the United States. Moreover, 5 anti-vaccine channels from the United States and 2 from other countries were taken down. No channels from Brazil were removed during the the discussed period.

As it can be seen, while it is true that a large number (approximately 16\%) of anti-vaccine videos were removed, most of the anti-vaccine videos were still available at the end of the tracking period, which illustrates why the measures taken by YouTube are said to be insufficient. It is important to note that most, if not all, pro-vaccine content removals happened due to the video creators themselves and not due to actions taken by YouTube. An even lower percentage of removals was found when looking at channels, where only 7\% of anti-vaccine channels were removed.

Another interesting aspect to consider is that the United States registered the largest absolute number and percentage of removed anti-vaccine videos, as 601 out of 1755 videos were removed(34\%). Brazil registered the lowest ratio of removals, with only 306 out of 3121 videos being removed(9.8\%). This could suggest that YouTube measures for filtering videos are more effective in some countries.

\section{Conclusion}
In the context of the COVID-19 pandemic, online social media platforms were important sources of information for many people. However, amidst the endless stream of content posted on those platforms, misinformation was widespread. As anti-vaccine content flooded the social networks, the efforts to prevent their spread grew, leading to different reactions from the various platforms. YouTube, the platform of interest for this study due to its large reach, dealt with this kind of content by the means of demonetizing~\cite{burki2020online}, raising awareness via information panels \footnote{https://support.google.com/youtube/answer/9004474?hl=en}, reducing recommendation for questionable videos, and content removal. 

Despite the aforementioned attempts to filter content and reduce disinformation, most comments and videos were still accessible at the end of the data collection period, with only about 16\% of anti-vaccine videos being taken down. For this reason, it became clear that it is important to study the difference between anti and pro-vaccine content to ascertain what differentiates those two categories and what are common characteristics among people who consume these kinds of videos, seeking to understand the dynamics of "comment sections".

By analyzing comments in both anti and pro-vaccine videos, we found that, on average, English speaking users engage more with anti-vaccine content. This is specially noticeable when looking at the removed videos, which are supposedly more controversial, as the user-user engagement on those is significantly higher than the average of the database. Additionally, we note that positive sentiments seem to be more popular in pro-vaccine "comment sections", while the same can be said for negative sentiments in anti-vaccine "comment sections". Moreover, during the study, it was possible to observe higher toxicity probability on anti-vaccine comments, corroborating to the observation that negative statements are more popular in this environment. This is further explored when looking at the similarities between videos and their comments, as it is observed that anti-vaccine videos are more similar to their "comment section" when they show negative emotions, with the opposite being true for pro-vaccine content.

In conclusion, in this study, the removal of videos by the YouTube platform was monitored and different methods were used to analyze anti and pro-vaccine "comment sections", finding key differences between them. Future work could explore similar concepts for different social network platforms such as Facebook, Instagram, and Twitter, seeking to ascertain whether the tendencies observed match those found on YouTube. Finally,  the removal of content on YouTube could be explored over a larger time frame with a larger database, in an attempt to better understand the effectiveness of the actions taken by the platform.

\begin{acks}
 The authors would like to thank the participants of the  project ``Measuring The Impact Of Covid Misinformation Influence Operations On Public Health Outcomes’’, with support from the Carnegie Endowment for International Peace’s Partnership for Countering Influence Operations, for valuable discussions and feedbacks during the year of 2021.

This work was funded in part by Brazilian agencies CNPq, CAPES and FAPEMIG, and by projects INCTCyber, MASWeb and CIIA-Saude.
\end{acks}

\bibliographystyle{ACM-Reference-Format}
\bibliography{bibliography}

\end{document}